# Theoretical proposal of a revolutionary water-splitting photocatalyst: The monolayer of boron phosphide

Tatsuo Suzuki (鈴木達夫)

*Tokyo Metropolitan College of Industrial Technology, 8-17-1, Minami-Senju, Arakawa-ku, Tokyo 116-8523, Japan*

ABSTRACT

Recently, hydrogen generation by water-splitting photocatalysts is attracting attention as a sustainable and clean energy resource. Photocatalytic hydrogen-generation systems are much simpler, cheaper, and easier to scale up than the coupled systems of electrolysis and solar cells, wind-power generation, etc. However, photocatalytic hydrogen generation is currently inefficient. This paper proposes the monolayer of boron phosphide as a stable highly-efficient water-splitting photocatalyst by high-precision density-functional theory calculations using a HSE06 functional with a solvent effect. The monolayer of boron phosphide has a direct allowed energy gap of about 1.4 eV, and functions as a one-step excitation photocatalyst. It absorbs sunlight with wavelengths below about 890 nm (ultraviolet, visible, and near-infrared light) and produces both hydrogen gas and oxygen gas from water at a suitable pH condition. By calculating the overpotentials of hydrogen and oxygen evolution reactions, its photocatalytic effectiveness was confirmed. The monolayers of boron phosphide will realize green hydrogen revolution.

## 1. Introduction

Fossil fuels cause global warming and will dry up in the future. Nuclear fuel has a risk of radioactive contamination such as Fukushima or Chernobyl. Recently, hydrogen generation by water-splitting photocatalysts [1] is attracting attention as a sustainable and clean energy resource to replace fossil fuels and nuclear fuel. Photocatalytic hydrogen-generation systems are much simpler, cheaper, and easier to scale up than the coupled systems of electrolysis and solar cells, wind-power generation, etc. Recently, Nishiyama et al. [2] reported solar hydrogen production from water on a 100 $m^2$-scale using an aluminum-doped strontium titanate particulate photocatalyst. They demonstrated that safe, large-sale photocatalytic water splitting and gas collection and separation were possible; however, the hydrogen production was inefficient. Finding a stable highly-efficient photocatalyst is an urgent task for humankind to realize green hydrogen revolution.

In 2012, Sun et al. [3] reported the fabrication of large-area free-standing single layers of ZnSe with four-atomic thickness, which have an enhanced water-splitting efficiency and photostability. The ZnSe single layers exhibit a photocurrent density of 2.14 mA $cm^{-2}$ at 0.72 V versus Ag/AgCl under 300 W Xe lamp irradiation, 195 times higher than that of bulk ZnSe. Similarly, the efficiency of $SnS_2$ single layers is 72 times higher than bulk $SnS_2$ [4], and SnS monolayers is 104 times higher than bulk SnS [5]. These reports reveal that two-dimensional materials can be highly-efficient water-splitting photocatalysts. The possible reason is as follows. In the case of bulk photocatalysts, light penetrates deep inside the photocatalyst, where it generates electron-hole pairs. The generated electrons and holes must move to the surface of the photocatalysts against the mutual Coulomb attraction in order to react with atoms or ions adsorbed on the surface. Therefore, bulk photocatalysts require mechanisms such as depletion layers or heterojunctions to separate electrons and holes; nevertheless, the electrons and holes still recombine, resulting in energy loss. However, in the case of two-dimensional photocatalysts, the generated electrons and holes can instantly react with atoms or ions adsorbed on the surface with little movement because two-dimensional photocatalysts are surface-only materials. Since Sun's reports, many two-dimensional photocatalysts have been developed, and details are found in review articles [6,7,8].

## 2. Proposal

This paper theoretically proposes the monolayer of boron phosphide (BP) as a stable highly-efficient water-splitting photocatalyst (cf. Fig. 1 (a)). As a result of searching for various two-dimensional materials by high-precision density-functional theory (DFT) calculations, the revolutionary property of the BP monolayer was discovered. At present,

*E-mail address:* dr.tatsuosuzuki@gmail.com.

there are no reports that BP monolayers were synthesized. However, in 2019, on the surface of a cubic zinc-blend BP nanocrystal grown at 1250 °C through a solid state reaction route, graphite-like layers with the lattice spacing of 0.35 nm were observed in high-resolution TEM images [9]. These layers may become the precursors of BP monolayers. In 2020, Hernández et al. [10] theoretically proposed that BP monolayers can be exfoliated by incorporating arsenic in the (1 1 1) surface of a cubic zinc-blend BP.

Below, we investigate the property of the BP monolayer as a photocatalyst by two types of high-precision DFT calculations: a plane-wave (PW) basis calculation using a pseudo-potential and a Gaussian type orbital (GTO) basis calculation using all electrons. The PW basis calculation also include a solvent effect, i.e., a water polarization effect by an implicit solvation model based on the Poisson-Boltzmann equation. All calculations use Heyd-Scuseria-Ernzerhof (HSE06) hybrid density functionals [11,12] for both structural optimizations and energy bands calculations because the HSE06 functional is one of the most reliable calculation methods, and the error between the calculated energy gaps and the experimental values is less than 10% [13].

## 3. Important conditions

First, we confirm four important conditions that two-dimensional photocatalysts must satisfy for large-scale practical use; (a) stable in water against long exposure to strong sunlight, (b) made from earth-abundant elements, (c) one-step excitation using a single semiconductor because Z-scheme using two connected semiconductors requires double photons of one-step excitation, and PEC water splitting using two electrodes is difficult to deploy on a large scale due to the complexity of equipment, and (d) a direct allowed transition semiconductor with an energy gap $E_g$ that is larger than the theoretical limit $\Delta_{limit}$ but as small as possible. In order to achieve one-step excitation of (c), band edges (i.e., conduction band minimum $E_C$ and valence band maximum $E_V$) should straddle water redox potentials $E_{H^+/H_2}$ and $E_{O_2/H_2O}$; that is,

$$E_C \geqslant E_{H^+/H_2} \quad \text{and} \quad E_{O_2/H_2O} \geqslant E_V \tag{1}$$

And, $\Delta_{limit}$ of (d) is the redox potential difference; that is,

$$E_g = E_C - E_V \geqslant \Delta_{limit} = E_{H^+/H_2} - E_{O_2/H_2O} \tag{2}$$

Next, we confirm redox potentials $E_{H^+/H_2}$ and $E_{O_2/H_2O}$. The hydrogen evolution reaction (HER) and the oxygen evolution reaction (OER) are expressed in Eqs. (3) and (4), respectively.

$$2\,H^+ + 2\,e^- \rightarrow H_2 \qquad (E^\circ = \ \ 0.000\ \text{V}\ vs.\ \text{SHE}) \tag{3}$$

$$2\,H_2O \rightarrow O_2 + 4\,H^+ + 4\,e^- \qquad (E^\circ = +1.229\ \text{V}\ vs.\ \text{SHE}) \tag{4}$$

Water molecules $H_2O$ and hydrogen ions $H^+$ adsorbed on the surface of the BP monolayer are transformed into various intermediates on the surface (cf. Fig. 1(b)–(h)), and are finally decomposed into hydrogen molecules $H_2$ and oxygen molecules $O_2$, as described in detail below. Here, $E^\circ$ is the standard electrode potential relative to the standard hydrogen electrode (SHE). $E_{H^+/H_2}$ and $E_{O_2/H_2O}$ are expressed by the Nernst equation in Eqs. (5) and (6), respectively.

$$E_{H^+/H_2} = -4.44 - \frac{k_B T}{2e} \ln\left(\frac{[H^+]^2}{p_{H_2}}\right) = -4.445 + 0.0592\,pH\ (\text{eV}) \tag{5}$$

$$E_{O_2/H_2O} = -4.44 - 1.229 - \frac{k_B T}{4e} \ln\left(p_{O_2} \cdot [H^+]^4\right) = -5.662 + 0.0592\,pH\ (\text{eV}) \tag{6}$$

From the definition of the Nernst equation, $E_{H^+/H_2}$ and $E_{O_2/H_2O}$ are the potential energies of the electron in the BP monolayer, not the potential energies in the solution. The origin of these equations is the vacuum level, and −4.44 eV is the energy level of SHE relative to the vacuum level. $k_B$, $e$, and $[H^+]$ are Boltzmann constant, elementary charge, and the molar concentration of $H^+$ ions, respectively. Temperature T = 298.15 K, and $pH = -\log_{10}[H^+]$. The reaction vessel is filled with generated gas; therefore, the partial pressures of $H_2$ and $O_2$ are $p_{H_2} = \frac{2}{3}$ atm and $p_{O_2} = \frac{1}{3}$ atm, respectively. The theoretical limit $\Delta_{limit} = E_{H^+/H_2} - E_{O_2/H_2O} = 1.217$ eV.

## 4. Reviews of previous researches

First, we review the bulk crystal of BP, which has a cubic zinc-blend structure. The cubic zinc-blend BP is quite chemically stable and is resistant to chemical corrosion. It is not attacked by hot concentrated mineral acids or aqueous alkali solution [14]. It is also resistant to oxidation in air up to about 800–1000 °C [15]. The cubic zinc-blend BP is an indirect transition semiconductor with an energy gap of 2.0 eV, and functions as a photocatalyst for $H_2$ evolution [9,14,16]. The photocatalytic $H_2$ evolution reactions continue even in strong acid or strong alkaline, and the BP photocatalyst is stable under these extreme conditions [14]. These durable properties of the cubic zinc-blend BP originate in the strong covalent bonds between elements B and P. This durability will be carried over to BP monolayers. Recently, Li et al. [17] prepared cubic zinc-blend BP nanosheets with a thickness of about 4 nm, and demonstrated that the BP nanosheets were well dispersed in water with a concentration of 0.2 mg/mL. Liang et al. [18] synthesized cubic zinc-blend BP single crystals larger than 1 mm, and evaluated their excellent thermal stability up to 1200 °C.

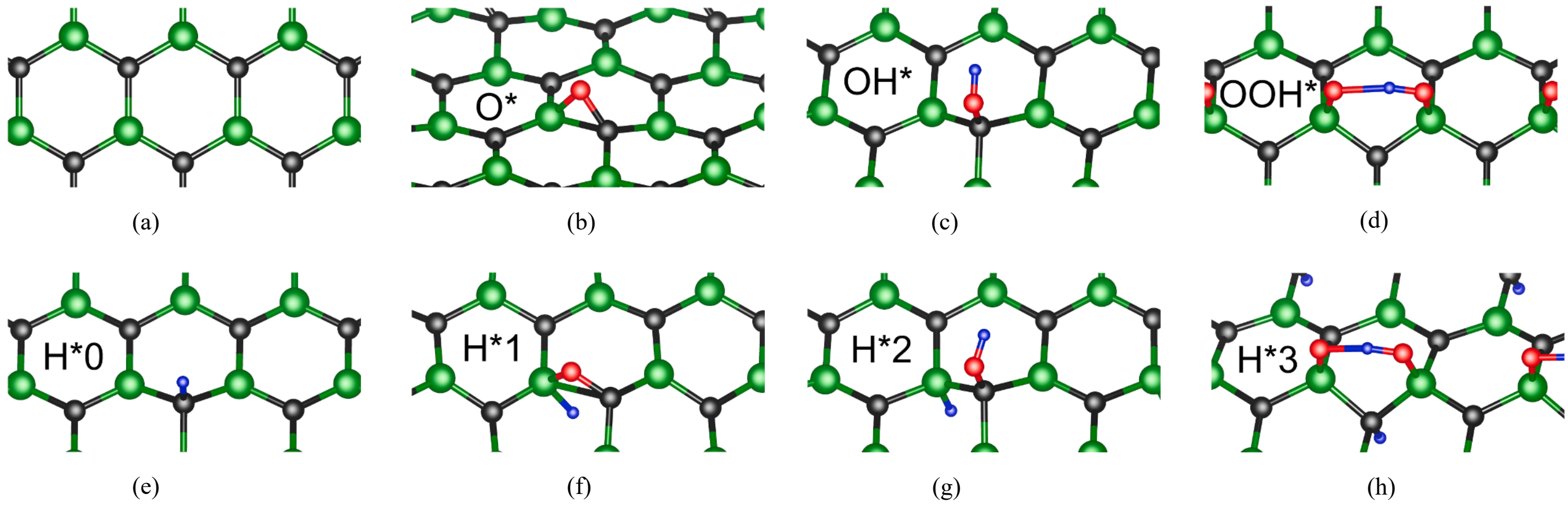

**Fig. 1.** Optimized structures of (a) BP monolayer, (b) O*, (c) OH*, (d) OOH*, (e) H*0, (f) H*1, (g) H*2, and (h) H*3. Black, green, red, and blue balls are boron, phosphorus, oxygen, and hydrogen atoms, respectively. (For interpretation of the references to colour in this figure legend, the reader is referred to the web version of this article.)

Next, we review BP monolayers. The most stable structure of a BP monolayer is a hexagonal planar structure like a graphene or a h-BN monolayer (cf. Fig. 1(a)) [19]. Şahin et al. [20] performed DFT calculations using a PW basis and local-density approximation, and reported the bond length $d = 1.83$ Å and the direct energy gap $E_g =$ 0.82 eV or 1.81 eV which is corrected by frequency-dependent $GW_0$ calculations. They calculated the phonon dispersion of a BP monolayer also, and confirmed the dynamical stability of a BP monolayer. Suzuki & Yokomizo [21] calculated $d = 1.87$ Å and $E_g = 1.912$ eV using a GTO basis and a B3LYP/6-31G(d) functional. After that, the high-precision calculation method using a HSE06 functional became widespread. Zhuang & Hennig [22] calculated $d = 1.855$ Å and $E_g = 1.36$ eV using a PW basis and a HSE06 functional. Wang & Wu [23] calculated the band edges of a BP monolayer in a vacuum using a PW basis and a HSE06 functional: $d =$ 1.855 Å, $E_C = -3.96$ eV, and $E_V = -5.33$ eV; and showed that the optical absorption of a BP monolayer was strong over a wide energy range between 1.37 and 4 eV. They performed a BOMD simulation by using the Nosé-Hoover method at 2500 K for 5 ps, and indicated the high thermal stability of BP monolayers. Furthermore, they reported the chemical stability of BP monolayers in environment, such as $N_2$, $O_2$, $H_2O$, $H_2$, and $CO_2$. However, they did not mention the photocatalytic application of BP monolayers. Shu et al. [24] stated that BP monolayers were suitable for a water-splitting photocatalyst based on their calculations using a PW basis, a PBE functional, and the extrapolation of $G_0W_0$ method. However, their low-precision calculation overestimated the energy gap as $E_g = 1.833$ eV. Therefore, the intrinsic high-efficiency of BP monolayers was not reported correctly, and the photocatalytic effectiveness of BP monolayers has been buried. There are no other papers reporting photocatalytic applications of BP monolayers, except for van der Waals heterostructures with other monolayers [25–28]. Wu et al. [29] performed DFT calculations combined with *ab initio* molecular dynamics (AIMD) simulations, and demonstrated that the physical and chemical stabilities of BP monolayers in a vacuum, and in oxygen, water, and oxygen-water environment.

## 5. Calculation methods

PW basis calculations use a plane-wave basis set, the projector augmented wave (PAW) potentials [30,31], a HSE06 functional, a 27 × 27 × 1 Monkhorst-Pack k-point mesh, the super-cell width of 18 Å, the cutoff energy of 800 eV for the PW basis, and VASP 5.4.1 package [32,33]. The convergence criteria of electronic self-consistent calculations and ionic relaxations are $10^{-6}$ and $10^{-5}$ eV, respectively. A solvent effect by the polarization of water is implemented as an implicit solvation model based on the Poisson-Boltzmann equation, and is performed by using VASPsol package [34,35]. GTO basis calculations use a Gaussian type orbital 6–311G(d,p) basis set and a HSE06 functional under the periodic boundary condition. A pseudo-potential is not used because of all-electron calculations. About 2000 k-points are requested in a unit cell. The convergences of electronic self-consistent calculations are that the maximum and the root mean square (RMS) of the variations of the density matrix are less than $10^{-5}$ and $10^{-7}$, respectively, and that the variation of the total-energy is less than $10^{-5}$ Hartree. The convergences of ionic relaxations are that the maximum and RMS of force are less than $2 \times 10^{-6}$ and $10^{-6}$ Hartree/Bohr, respectively, and that the maximum and RMS of displacement are less than $6 \times 10^{-6}$ and $4 \times 10^{-6}$ Bohr, respectively. The GTO basis calculation is performed by using Gaussian 09 package [36].

## 6. Results and considerations

This paper shows three types of calculated results: (a) **PWsol** using a PW basis with a solvent effect, (b) **PWvac** using a PW basis in a vacuum, and (c) **GTOvac** using a GTO basis in a vacuum.

First, we consider the shapes of BP monolayers after structural optimizations. The bond lengths of PWsol, PWvac, and GTOvac are 1.845, 1.846, and 1.854 Å, respectively. Comparing PWsol and PWvac, the bond lengths are almost the same with and without a solvent effect. Comparing PWvac and GTOvac, there is 0.4% difference in bond lengths due to the different calculation methods. These bond lengths are in good agreement with previous researches [22,23].

Next, we consider the results of PWsol in Fig. 2. An infinitely wide BP monolayer is placed on the x-y plane at z = 0. $V(z)$ is the potential which is averaged within the unit cell.

$$V(z) = \frac{\iint_{\text{unit cell}} V_{\text{sol}}(x, y, z)\ dx\ dy}{\iint_{\text{unit cell}} dx\ dy} \tag{7}$$

where $V_{\text{sol}}(x,y,z)$ is the entire local potential (ionic plus Hartree plus exchange correlation) with the solvent effect. The potential value at the midpoint of the spacer is vacuum level, and it is aligned with the origin of energy; that is, $V(z = 9) = 0$. $n(z)$ is the averaged number density of valence electrons of the BP monolayer with the solvent effect, and $n_b(z)$ is the averaged bound charge density of water. Drawn for comparison, $V_{\text{vac}}(z)$ is the averaged potential of PWvac, i.e., the averaged potential without the solvent effect. $V_{\text{vac}}(z)$ at the midpoint of the spacer is also aligned with the origin of energy; that is, $V_{\text{vac}}(z = 9) = 0$. We define $V_d(z) \equiv V(z) - V_{\text{vac}}(z)$, which is the potential difference caused by the polarization of water. Here, we notice that $V_d(z)$ is the vacuum level when the solvent effect exists; that is, $V_d(z)$ corresponds to the vacuum level of Anderson model [37] which explains the energy band profiles of semiconductor heterostructures. Therefore, $V_H \equiv V_d(z = 0) = 0.21$ eV is the electrical double layer voltage. $E_C = -3.85$ eV and $E_V = -5.17$ eV are band edges with the solvent effect. $E_{H^+/H_2} = -3.90$eV and $E_{O_2/H_2O} = -5.12$eV are the redox potentials in Eqs. (5) and (6) at $pH =$ 9.2. Here, $E_{H^+/H_2}$ and $E_{O_2/H_2O}$ are based on the vacuum level $V_d(z > 3)$ outside the electric double layer. Under the condition in this figure, the band edges straddle the redox potentials, and Eqs. (1) and (2) are satisfied.

Then, we consider energy bands in the left panel of Fig. 3. Here, the energy bands of PWsol and PWvac are drawn under the same condition that $V(z = 9) = V_{\text{vac}}(z = 9) = 0$, as in Fig. 2. The energy bands of GTOvac are also drawn so that the vacuum level is the origin of energy. First, we compare PWsol and PWvac. The energy gaps of PWsol and PWvac are 1.32 and 1.34 eV, respectively. We understand that the difference of energy gaps between with and without the solvent effect is small. The energy gap of PWvac is in good agreement with previous researches [22,23]. The band edges $(E_C, E_V)$ of PWsol and PWvac are $(-3.85, -5.17)$ and $(-4.06, -5.39)$ eV, respectively. $(E_C, E_V)$ are shifted only by the electrical double layer voltage $V_H = 0.21$ eV. Next, we compare PWvac and GTOvac. The energy gap of GTOvac is 1.47 eV, which is 10% larger than PWvac. This discrepancy is due only to the different calculation methods. $(E_C, E_V)$ of GTOvac is $(-4.13, -5.60)$ eV. Despite the different calculation methods, the band edges $(E_C, E_V)$ are located at almost the same energy levels, which are in good agreement with a previous research [23].

Then, we consider the relationship between band edges and redox potentials. The band edges in this paper are calculated under the condition that the BP monolayer is not charged, i.e., the pH of the solution is at *the point of zero charge* $pH_{pzc}$. It is unclear how the band edges of the BP monolayer depend on pH; that is, whether the band edges show a Nernstian dependence of 0.059 eV/pH like semiconducting metal oxides (e.g. $TiO_2$) [38], or are pH-independent like a hydrogen-terminated diamond [39]. Therefore, we cannot estimate $pH_{pzc}$. In the right panel of Fig. 3, the redox potentials are drawn according to Eqs. (5) and (6). If the $pH_{pzc}$ of PWsol is in the range of 8.4–10.0, the BP monolayer functions as a photocatalyst because it satisfies Eq. (1). However, even if the $pH_{pzc}$ is out of this range, the BP monolayer still functions as a photocatalyst. This is because the band edges $E_C$ and $E_V$ are very close to the redox potentials $E_{H^+/H_2}$ and $E_{O_2/H_2O}$, respectively; then, either band edge is pinned to the corresponding redox potential. The basis for this idea is the experiments by Chakrapani et al. [39]. They added hydrogen-terminated diamond particles into aqueous solutions and measured the changes in pH and oxygen concentrations. Their experiments show that electron exchange systematically occurs between diamond and the oxygen redox couple; that is, the electrochemical potential (Fermi energy) of the diamond is pinned by the oxygen redox potential. By the way, on the BP monolayer side of water–solid interface there is little charge to compensate for ions adsorbed on the water side because the BP monolayer does not have free electrons like metals nor does it form a depletion layer like bulk semiconductors. Furthermore, the covalent bonds between elements B and P are so strong that B or P does not dissolve as ions. Therefore, the BP monolayer will not be charged very much.

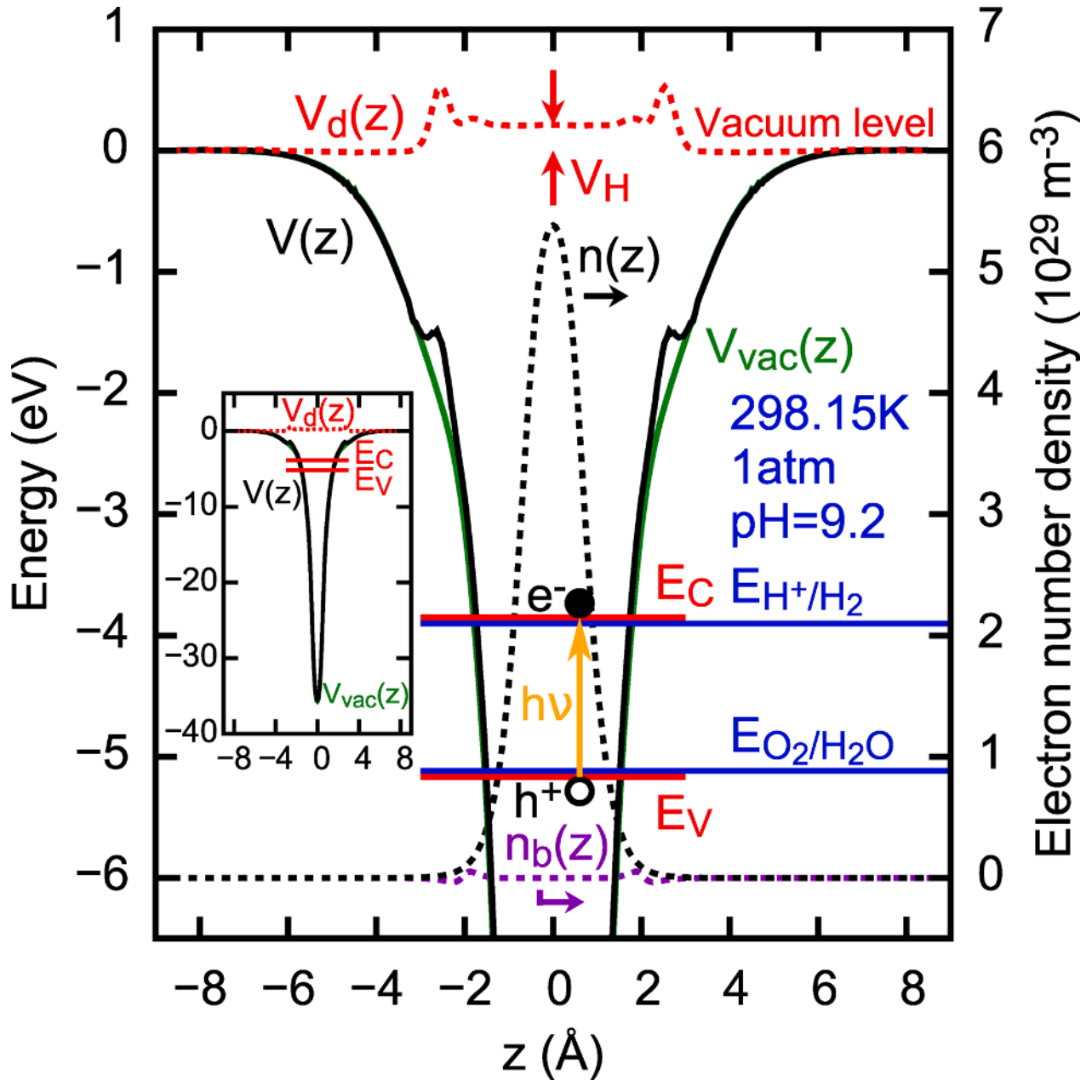

**Fig. 2.** Potentials, charge densities, and band edges vs. a z-coordinate. The inset is an overall view.

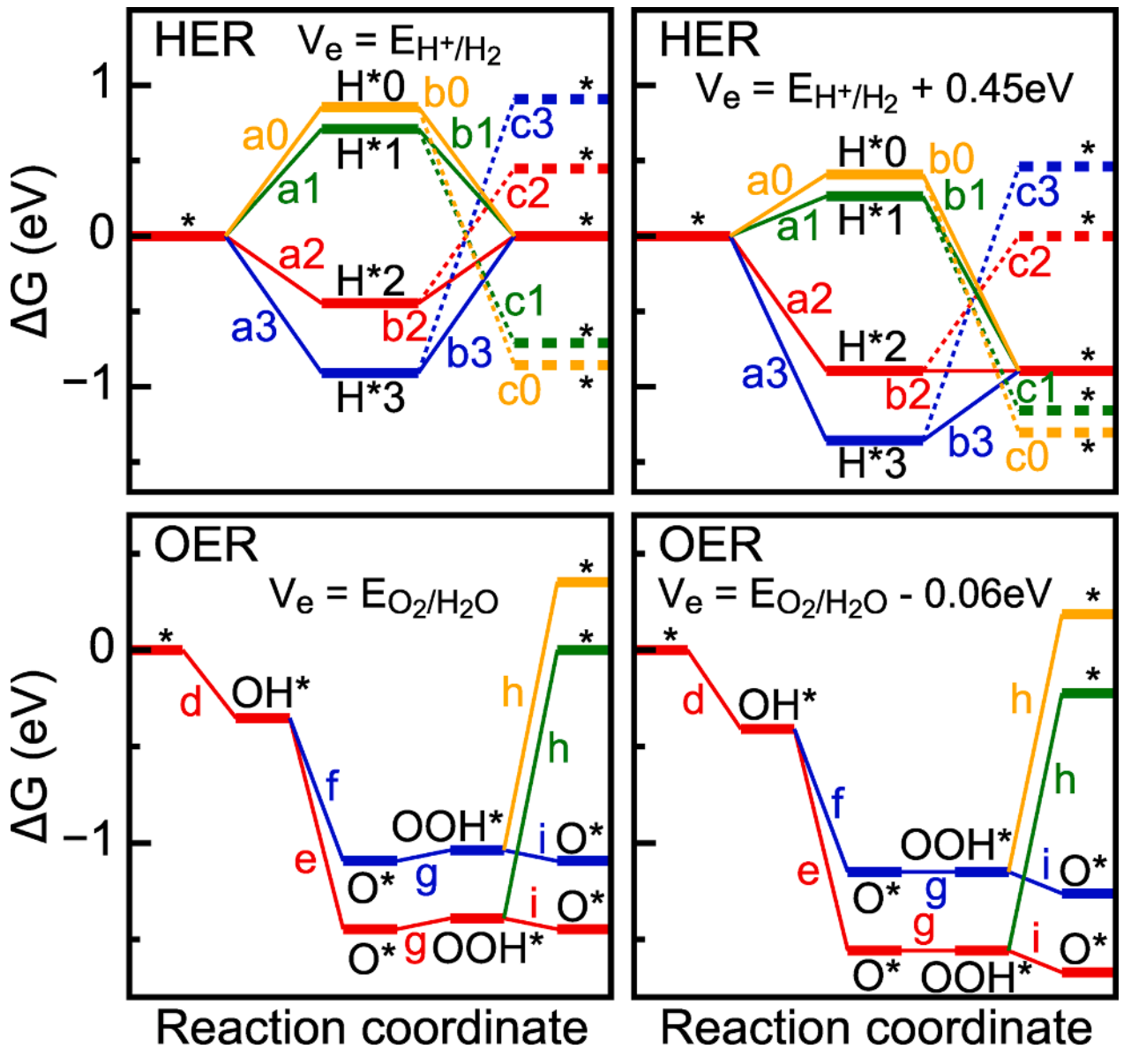

**Fig. 4.** Gibbs free energy change $\Delta G$ in each pathway. Here, **a0** means **a** reaction with a **H*0** adsorbed-proton. **a1**, **a2**, …, **c3** are similar.

Then, we calculate the overpotentials of HER and OER by Nørskov approach [40,41,42]. Here, we consider an acidic reaction involving $H^+$; however, the results are the same for the alkaline reaction involving $OH^-$ [43]. HER pathways considered are **(a)** $H^+ + e^- + * \rightarrow H*$, **(b)** $H^+ + e^- + H* \rightarrow H_2(g) + *$, and **(c)** $2H* \rightarrow H_2(g) + 2\,*$. OER pathways are **(d)** $H_2O(l) + * \rightarrow OH* + H^+ + e^-$, **(e)** $OH* \rightarrow O* + H^+ + e^-$, **(f)** $2OH* \rightarrow H_2O(l) + O* + *$, **(g)** $H_2O(l) + O* \rightarrow OOH* + H^+ + e^-$, **(h)** $OOH* \rightarrow O_2(g) + * + H^+ + e^-$, and **(i)** $2OOH* \rightarrow O_2(g) + 2O* + 2H^+ + 2\,e^-$. Here, * denotes an adsorb-site on the BP monolayer. In the following, temperature T = 298.15 K. The Gibbs free energies per molecule of hydrogen gas and oxygen gas are $G_{H_2(g)} = G^0_{H_2(g)} + \frac{k_B T}{e}\ln p_{H_2}$ and $G_{O_2(g)} = G^0_{O_2(g)} + \frac{k_B T}{e}\ln p_{O_2}$, where $G^0$ is the Gibbs free energy at standard conditions. The Gibbs free energies of water is $G_{H_2O(l)} = G^0_{H_2O(l)} = G^0_{H_2O(g)} - 0.088$ eV [44]. ${G^0}_{H_2(g)} = E^{DFT}_{H_2(g)} + E^{ZPE}_{H_2(g)} + \Delta {H^0}_{H_2(g)}(0\rightarrow T) - TS^0_{H_2(g)}$, and ${G^0}_{H_2O(g)} =$ $E^{DFT}_{H_2O(g)} + E^{ZPE}_{H_2O(g)} + \Delta H^0_{H_2O(g)}(0 \rightarrow T) - TS^0_{H_2O(g)}$. Here, $E^{DFT}_{H_2(g)}$ and $E^{DFT}_{H_2O(g)}$ are the total energies calculated by DFT. $E^{ZPE}_{H_2(g)} = \frac{1}{2e} h\nu_{H_2(g),1}$ and $E^{ZPE}_{H_2O(g)} = \frac{1}{2e}\sum_{i=1}^{3} h\nu_{H_2O(g),i}$ are the zero-point energies; where $h$ is Plank's constant, and $\nu_{H_2(g),1}$ and $\nu_{H_2O(g),\,i}$ are the calculated frequencies of the normal mode. The enthalpy changes $\Delta H^0{}_{H_2(g)}(0K \rightarrow T)$, $\Delta H^0{}_{H_2O(g)}(0K \rightarrow T)$ and the standard entropies $S^0_{H_2(g)}$, $S^0_{H_2O(g)}$ are taken from the Ref. [44]. $G^0_{O_2(g)} = 2G^0_{H_2O(l)} - 2G^0_{H_2(g)} + 4 \times 1.229$ eV which is E° of OER. The sum of Gibbs free energies of a proton and an electron is $G_{H^+} + G_{e^-} = \frac{1}{2}\ G^0_{H_2(g)} - \frac{k_BT}{e}\ln 10 \cdot pH + V_e + 4.44$ eV; where $V_e$ is the potential energy of the electron in the BP monolayer relative to the vacuum level, and 4.44 eV is the difference between SHE and the vacuum level. The Gibbs free energy of O* is calculated by $G_{O^*} = E^{DFT}_{O^*} + E^{ZPE}_{O^*} + \Delta U^0_{O^*}(0 \rightarrow T) - TS^0_{O^*} - E^{DFT}_{*}$. Here $E^{DFT}_{O^*}$ and $E^{DFT}_{*}$ are the total energies by DFT. $E^{ZPE}_{O^*}$, $\Delta U^0_{O^*}(0 \rightarrow T)$, and $S^0_{O^*}$ are the zero-point energy, the internal energy change, and the standard entropy, respectively. $E^{ZPE}_{O^*} = \frac{1}{2e}\sum_i h\nu_{O^*,i}$, and $\Delta U^0_{O^*}(0 \rightarrow T) - TS^0_{O^*} = \frac{k_BT}{e}\sum_i \ln\left\{1 - \exp\left(-\frac{h\nu_{O^*,i}}{k_BT}\right)\right\}$. Similarly, $G_{OH^*}$, $G_{OOH^*}$, and $G_{H^*}$ are calculated; however, for $G_{H^*}$ we consider four types: $G_{H^*0}$, $G_{H^*1}$, $G_{H^*2}$, and $G_{H^*3}$. H*0 is a proton adsorbed on the pure surface of the BP monolayer. H*1, H*2, and H*3 are protons adsorbed on the back surface of O*, OH*, and OOH*, respectively (cf. Fig. 1). Spin-polarized calculations were performed by using VASP 5.4.4 package, PAW potentials, a PBE functional [45], a DFT-D3 method for vdW interactions [46], and the cutoff energy of 520 eV for the PW basis. The convergence criteria of self-consistent calculations and ionic relaxations were $10^{-5}$ and $10^{-4}$ eV, respectively. The adsorbed atoms were calculated with a 2 × 2 × 1 supercell, a 7 × 7 × 1 Monkhorst-Pack k-point mesh, and the super-cell width of 18 Å. $H_2$(g) and $H_2O$(g) were calculated in a box of 20 Å×20 Å×20 Å at Γ k-point. As a result, $G_{H_2(g)}$, $G_{O_2(g)}$, $G_{H_2O(l)}$, $G_{O^*}$, $G_{OH^*}$, $G_{OOH^*}$, $G_{H^*0}$, $G_{H^*1}$, $G_{H^*2}$, $G_{H^*3}$ and $G_{H^+} + G_{e^-}$ are −6.827, −9.911, −14.216, −6.403, −9.938, −15.932, −2.557, −2.703, −3.862, −4.324, and $1.032 - 0.0592\ pH + V_e$ eV, respectively (see the supplementary information for computational details). For each reaction (a)–(i), we calculate the Gibbs free energy change $\Delta G$. Because the reaction does not proceed unless $\Delta G < 0$, we find the threshold potential $V_{e,th}$ when $\Delta G = 0$. The difference between this threshold potential and the redox potential is defined as $\eta$; that is, $\eta = V_{e,th} - E_{H^+/H_2}$ for HER and $\eta = E_{O_2/H_2O} - V_{e,th}$ for OER. Note that $\eta$ is pH independent because the threshold potential and the redox potential are the same pH dependent. For all reactions included in a pathway, the maximum value of those $\eta$ is the overpotential $\eta_{HER}$ or $\eta_{OER}$. Fig. 4 shows Gibbs free energy change $\Delta G$ in each pathway. Therefore, $\eta_{HER} = 0.45$ eV along a pathway (a and b) with a H*2 adsorbed-proton; and $\eta_{OER} = 0.06$ eV along a pathway (d, e, g, and i) or (d, f, g, and i). By the way, the difference between the energy gap and the theoretical limit $E_g - \Delta_{limit}$ are 0.10, 0.12, and 0.25 eV for PWsol, PWvac, and GTOvac, respectively. Although these values are a little smaller than $\eta_{HER} + \eta_{OER}$, the BP monolayer will function well as a photocatalyst under sunlight. The reason is as follows. When a photon with an energy larger than the energy gap is absorbed, an electron with an energy larger than $E_C$ and a hole with an energy smaller than $E_V$ are generated. Before the electron falls to the bottom of the conduction band, the hole rises to the top of the valence band, or the electron and the hole recombine or form an exciton, the electron and the hole can react with atoms or ions adsorbed on the surface of the BP monolayer because the BP monolayer is a surface-only material. This is a great feature that is different from bulk photocatalysts.

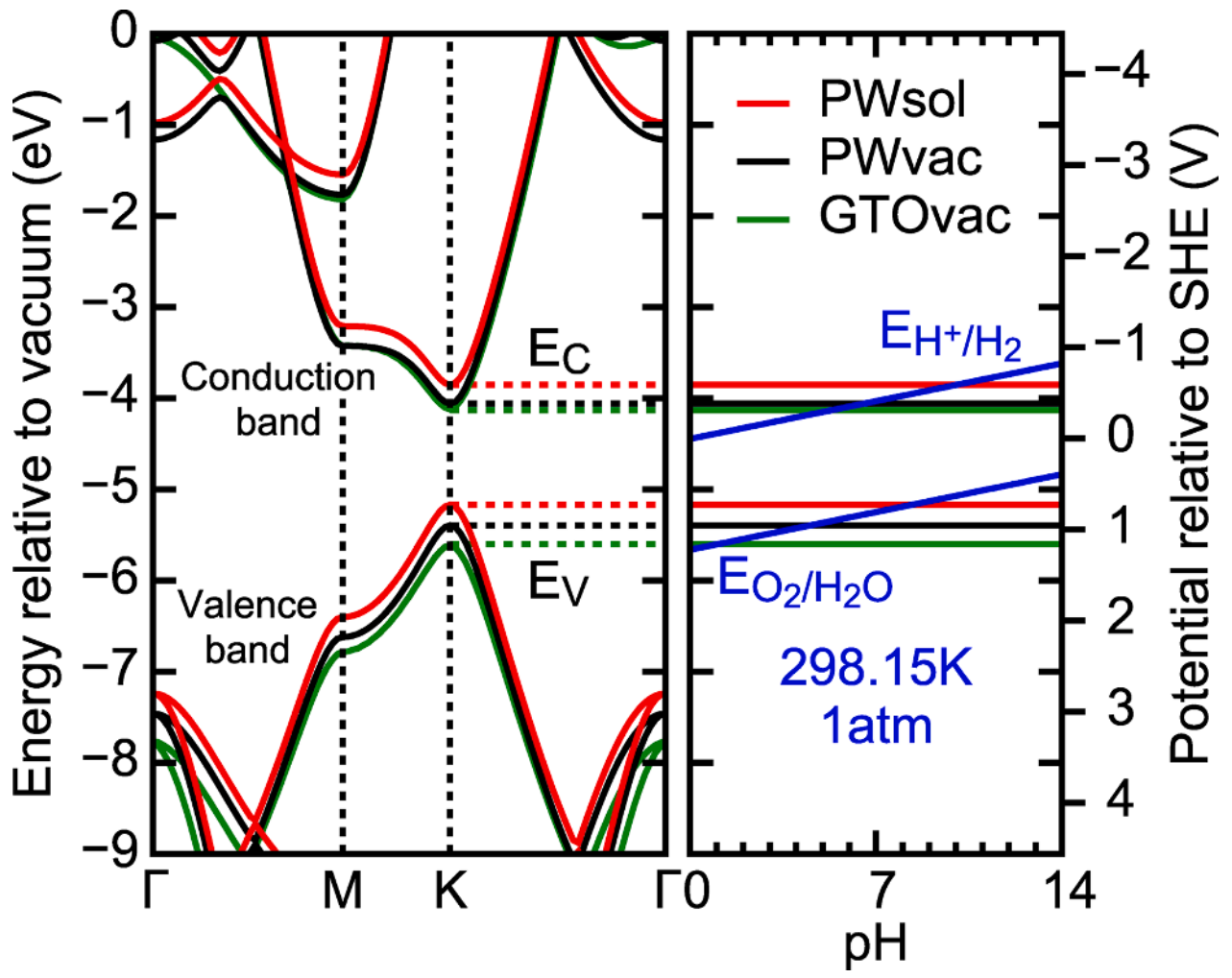

**Fig. 3.** (Left panel) energy bands. (Right panel) band edges and redox potentials vs. pH.

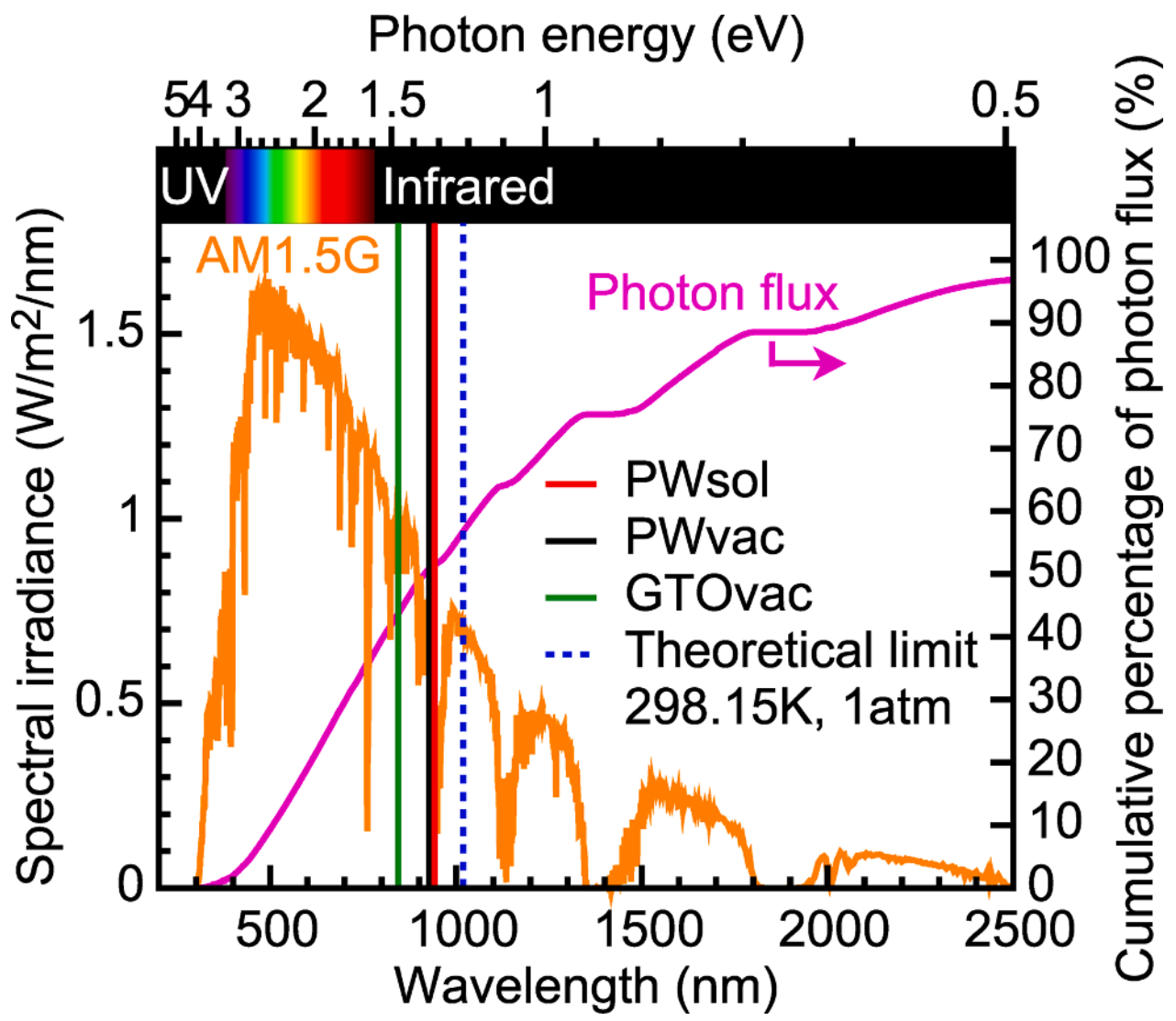

**Fig. 5.** Calculated energy gaps, solar radiation spectrum (air mass 1.5 global), and photon flux vs. wavelength.

And finally, Fig. 5 shows the calculated energy gaps of BP monolayers and solar radiation spectrum (air mass 1.5 global) [47]. The BP monolayer has an energy gap of about 1.4 eV. It absorbs sunlight with wavelengths below about 890 nm (ultraviolet, visible, and near-infrared light), and uses about 48% of the photon flux from the sun effectively.

## 7. Conclusions

This paper theoretically proposes the BP monolayer as a highly-efficient water-splitting photocatalyst. It is a stable semiconductor with a direct allowed energy gap of about 1.4 eV, and functions as a one-step excitation photocatalyst. It absorbs sunlight with wavelengths below about 890 nm (ultraviolet, visible, and near-infrared light) and produces both $H_2$ and $O_2$ from water at a suitable pH condition. By calculating the overpotentials of hydrogen and oxygen evolution reactions, its photocatalytic effectiveness was confirmed. BP monolayers will realize the hydrogen economy as a sustainable and clean energy resource; therefore, we hope that BP monolayers will be synthesized.

## Declaration of Competing Interest

The author declares that he has no known competing financial interests or personal relationships that could have appeared to influence the work reported in this paper.

## Acknowledgements

This work was partially supported by Tokyo Metropolitan College of Industrial Technology.

## Appendix A. Supplementary material

Supplementary data to this article can be found online at https://doi.org/10.1016/j.apsusc.2022.153844.

# Supplementary Information

Theoretical proposal of a revolutionary water-splitting photocatalyst: The monolayer of boron phosphide

Tatsuo Suzuki (鈴木 達夫)

Tokyo Metropolitan College of Industrial Technology

**TABLE I.** Calculated results: Gibbs free energy $G$, total energy calculated by DFT $E^{\mathrm{DFT}}$, zero-point energy calculated by DFT $E^{\mathrm{ZPE}}$, internal energy change $\Delta U^0(0\mathrm{K} \rightarrow T)$, enthalpy change $\Delta H^0(0\mathrm{K} \rightarrow T)$, standard entropy $S^0$, and Gibbs free energy at standard conditions $G^0$. The unit is eV.

| | $G$ | $E^{\mathrm{DFT}}$ | $E^{\mathrm{ZPE}}$ | $\Delta U^0(0\mathrm{K} \rightarrow T) - TS^0$ or $\Delta H^0(0\mathrm{K} \rightarrow T) - TS^0$ | $G^0$ |
|---|---|---|---|---|---|
| * | | -49.059 | | | |
| O* | -6.403 | -55.545 | 0.094 | -0.011 | |
| OH* | -9.938 | -59.335 | 0.374 | -0.036 | |
| OOH* | -15.932 | -65.375 | 0.439 | -0.055 | |
| H*0 | -2.557 | -51.850 | 0.235 | -0.001 | |
| H*1 | -2.703 | -58.494 | 0.247 | -0.001 | |
| H*2 | -3.862 | -63.427 | 0.232 | -0.002 | |
| H*3 | -4.324 | -69.940 | 0.243 | -0.001 | |
| $H_2$(g) | -6.827 | -6.771 | 0.271 | -0.316 | -6.816 |
| $H_2O$(g) | | -14.219 | 0.572 | -0.481 | -14.128 |
| $H_2O$(l) | -14.216 | | | | -14.216 |
| $O_2$(g) | -9.911 | | | | -9.883 |
| $H^+ + e^-$ | $1.032 - 0.0592\,pH + V_e$ | | | | |

**TABLE II.** Gibbs free energy change $\Delta G$ and the difference $\eta$ between the threshold potential and the redox potential. The unit is eV.

**HER pathways**

(a0) $H^+ + e^- + * \rightarrow H*0$ $\quad \Delta G = -3.589 + 0.0592\,pH - V_\mathrm{e}$ $\quad \eta = 0.856$
(b0) $H^+ + e^- + H*0 \rightarrow H_2(g) + *$ $\quad \Delta G = -5.302 + 0.0592\,pH - V_\mathrm{e}$ $\quad \eta = -0.856$
(c0) $2\ H*0 \rightarrow H_2(g) + 2\ *$. $\quad \Delta G = -1.713$

(a1) $H^+ + e^- + * \rightarrow H*1$ $\quad \Delta G = -3.735 + 0.0592\,pH - V_\mathrm{e}$ $\quad \eta = 0.711$
(b1) $H^+ + e^- + H*1 \rightarrow H_2(g) + *$ $\quad \Delta G = -5.156 + 0.0592\,pH - V_\mathrm{e}$ $\quad \eta = -0.711$
(c1) $2\ H*1 \rightarrow H_2(g) + 2\ *$. $\quad \Delta G = -1.421$

(a2) $H^+ + e^- + * \rightarrow H*2$ $\quad \Delta G = -4.894 + 0.0592\,pH - V_\mathrm{e}$ $\quad \eta = -0.448$
(b2) $H^+ + e^- + H*2 \rightarrow H_2(g) + *$ $\quad \Delta G = -3.997 + 0.0592\,pH - V_\mathrm{e}$ $\quad \eta = 0.448$
(c2) $2\ H*2 \rightarrow H_2(g) + 2\ *$. $\quad \Delta G = 0.897$

(a3) $H^+ + e^- + * \rightarrow H*3$ $\quad \Delta G = -5.356 + 0.0592\,pH - V_\mathrm{e}$ $\quad \eta = -0.910$
(b3) $H^+ + e^- + H*3 \rightarrow H_2(g) + *$ $\quad \Delta G = -3.535 + 0.0592\,pH - V_\mathrm{e}$ $\quad \eta = 0.910$
(c3) $2\ H*3 \rightarrow H_2(g) + 2\ *$. $\quad \Delta G = 1.821$

**OER pathways**

(d) $H_2O(l) + * \rightarrow OH* + H^+ + e^-$ $\quad \Delta G = 5.309 - 0.0592\,pH + V_\mathrm{e}$ $\quad \eta = -0.353$

(e) $OH* \rightarrow O* + H^+ + e^-$ $\quad \Delta G = 4.568 - 0.0592\,pH + V_\mathrm{e}$ $\quad \eta = -1.094$

(f) $2\ OH* \rightarrow H_2O(l) + O* + *$ $\quad \Delta G = -0.741$

(g) $H_2O(l) + O* \rightarrow OOH* + H^+ + e^-$ $\quad \Delta G = 5.718 - 0.0592\,pH + V_\mathrm{e}$ $\quad \eta = 0.056$

(h) $OOH* \rightarrow O_2(g) + * + H^+ + e^-$ $\quad \Delta G = 7.053 - 0.0592\,pH + V_\mathrm{e}$ $\quad \eta = 1.391$

(i) $2\ OOH* \rightarrow O_2(g) + 2\ O* + 2\ H^+ + 2\ e^-$
$\quad \Delta G = 11.211 + 2(-0.0592\,pH + V_\mathrm{e})$ $\quad \eta = -0.056$